\documentclass[10pt,journal,compsoc]{IEEEtran}
\ifCLASSOPTIONcompsoc
  \usepackage[nocompress]{cite}
\else
  \usepackage{cite}
\fi
\ifCLASSINFOpdf
   \usepackage[pdftex]{graphicx}
   \graphicspath{{images/}}
  \DeclareGraphicsExtensions{.pdf,.png,.jpg}
\else
\fi
\usepackage{amsmath}
\usepackage{amssymb}
\usepackage{mathtools}
\usepackage{pifont}

\usepackage{algorithmic}
\usepackage{listings}
\usepackage[dvipsnames]{xcolor}
\usepackage[scaled=0.85]{beramono} 
\usepackage{array}
\usepackage{diagbox}

\ifCLASSOPTIONcompsoc
  \usepackage[caption=false,font=footnotesize,labelfont=sf,textfont=sf]{subfig}
\else
  \usepackage[caption=false,font=footnotesize]{subfig}
\fi
\usepackage{fixltx2e}

\usepackage{stfloats}
\usepackage{url}

\usepackage[english]{babel}

\begin{document}
{\large \textbf{IEEE Copyright Notice}}

© 2020 IEEE. Personal use of this material is permitted. Permission
from IEEE must be obtained for all other uses, in any current or future
media, including reprinting/republishing this material for advertising or
promotional purposes, creating new collective works, for resale or
redistribution to servers or lists, or reuse of any copyrighted
component of this work in other works.
\newpage

\title{Formal Verification of a Fail-Operational Automotive Driving System}
%
%
%
%
\author{Tobias~Schmid, Stefanie~Schraufstetter, Jonas~Fritzsch, Dominik~Hellhake, Greta~Koelln~and~Stefan~Wagner
\IEEEcompsocitemizethanks{\IEEEcompsocthanksitem T.Schmid, S.Schraufstetter and D.Hellhake are with the Department of Development of Driving Dynamics, BMW AG, Munich, Germany.\protect\\
\IEEEcompsocthanksitem G.Koelln is with the Department of Development of Autonomous Driving, BMW AG, Munich, Germany.\protect\\
\IEEEcompsocthanksitem T.Schmid, J. Fritzsch, D. Hellhake and S. Wagner are with the Institute of Software Engineering, University of Stuttgart, Stuttgart, Germany.}
\thanks{Manuscript received July 19, 2020; revised August 26, 2020.}}
%
%

\markboth{Journal of \LaTeX\ Class Files,~Vol.~14, No.~8, August~2015}%
{Shell \MakeLowercase{\textit{et al.}}: Bare Demo of IEEEtran.cls for Computer Society Journals}
%



\IEEEtitleabstractindextext{%
\begin{abstract}
A fail-operational system for highly automated driving must complete the driving task even in the presence of a failure. This requires redundant architectures and a mechanism to reconfigure the system in case of a failure. Therefore, an arbitration logic is used. For functional safety, the switch-over to a fall-back level must be conducted in the presence of any electric and electronic failure. To provide evidence for a safety argumentation in compliance with ISO 26262, verification of the arbitration logic is necessary. The verification process provides confirmation of the correct failure reactions and that no unintended system states are attainable. Conventional safety analyses, such as the failure mode and effect analysis, have its limits in this regard. We present an analytical approach based on formal verification, in particular model checking, to verify the fail-operational behaviour of a driving system. For that reason, we model the system behaviour and the relevant architecture and formally specify the safety requirements. The scope of the analysis is defined according to the requirements of ISO 26262. We verify a fail-operational arbitration logic for highly automated driving in compliance with the industry standard. Our results show that formal methods for safety evaluation in automotive fail-operational driving systems can be successfully applied. We were able to detect failures, which would have been overlooked by other analyses and thus contribute to the development of safety critical functions.
\end{abstract}

\begin{IEEEkeywords}
Automotive, highly automated driving, fail-operational, Fault-tolerance, Functional Safety, Dependability,  ISO 26262, Formal Verification, Model Checking
\end{IEEEkeywords}}

\maketitle

\IEEEdisplaynontitleabstractindextext

%
\IEEEpeerreviewmaketitle

\begin{figure}[t!]
\centering
\includegraphics[trim={0mm 0mm 0mm 0mm},clip,scale=0.4]{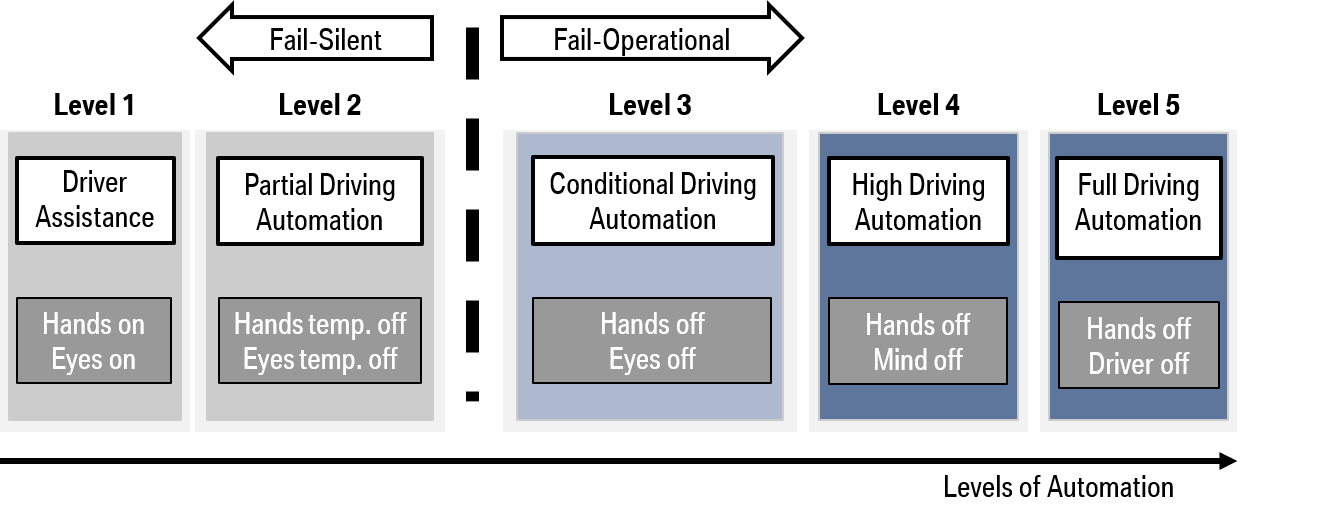} 
\caption{Levels of Driving Automation \cite{SAE}}
\label{fig:Levels_Automation}
\end{figure}

\IEEEraisesectionheading{\section{Introduction}\label{sec:introduction}}
\IEEEPARstart{I}{n} partial driving automation, the driver supervises the driving system at all times  (shown in figure \ref{fig:Levels_Automation}) and is therefore able to act as a backup in case of a failure. A failure in this context is the consequence of a fault leading to the loss of an element, for example, a steering system. The system’s failure reaction in partial driving automation is defined as fail-silent. At higher levels of automation, the driver does not necessarily intervene immediately, since s/he is not in charge of supervising the system permanently \cite{SAE}. Therefore, a highly automated driving or conditional driving system is required to operate even in case of a failure and initiates the transition to a safe state to attain functional safety. Those systems are defined as fail-operational. Functional safety is the absence of unreasonable risk due to malfunctions of electric and electronic (E/E) systems and is mandatory for the homologation of vehicle systems \cite{Winner}. The standard for functional safety in the automotive industry is ISO 26262 \cite{ISO26262}, which includes guidelines for the evaluation, management, and development of safety critical systems. 

The fail-operational behaviour of the driving system allows the availability of driving functions to prevent a hazardous situation.  Since every single E/E failure must be tolerated, redundant architectures are inevitable. In the case of the driving system, this includes redundant steering and braking systems as well as sensors and control algorithms. When a failure occurs, a reconfiguration and activation of a backup operation are necessary. The switch-over is ensured by an implemented arbitration logic.

\subsection{Problem Statement} 
The industry standard for functional safety, ISO 26262 \cite{ISO26262}, requires safety argumentation to evaluate the absence of unreasonable risk due to malfunctions of an E/E system. Verification provides evidence for the argumentation and the appropriateness of the safety concept. The verification analysis process verifies that the defined specifications and design meet higher-level requirements, such as system-level requirements. In the context of the fail-operational system behaviour it specifically includes arbitration logic, a logic to ensure the switch-over to a backup level. Established methods in the automotive industry, such as failure mode and effect analysis (FMEA), do not meet the challenges of such systems because a large number of states and propagation paths limit its practical application. To date, there is no published approach for a safety argumentation or verification process of such systems. 

\subsection{Research Objective}
The objective of our research is the verification of the fail-operational behaviour of a driving system and in particular, the arbitration logic in accordance with ISO 26262. Our verification approach should be reliable to contribute to an automotive safety case. In addition, the objective is to demonstrate the application of formal methods, in particular model checking for a safety argumentation of an industry-relevant, complex problem, a fail-operational driving system. The scope of our investigation especially includes the arbitration logic, and respective hardware interfaces such as power supply.

\subsection{Contribution}
Our approach demonstrates that model checking is a reliable analytical technique for the verification of a fail-operational driving system in compliance with the industry standard for functional safety, ISO 26262. The model integrates the arbitration logic and the relevant architecture using an open-source tool (NuSMV\footnote{\url{http://nusmv.fbk.eu/}}) and displays the transformation of the safety requirements into a formal specification using linear time logic. Furthermore, the requirements of the industry standard ISO 26262 regarding the scope and the verification process are identified. Relevant failure cases are derived from the architecture and considered in the model. We overcome the state space explosion problem during formal verification, by limiting the failure combinations to the scope of ISO 26262, segmenting the checking procedure, and bounding the search depth. Then the validation of the model and the formal specifications is conducted. In addition, a tool qualification is necessary for compliance with ISO 26262. The qualification ensures the reliability of the tool and therefore the resilience of the generated results. Our study demonstrates that verification via model checking is applicable for sophisticated and extensive automotive problems and it provides verification of accurateness of the system design in compliance with ISO 26262. Thus, this study develops a safety argumentation of a fail-operational driving system in the development of highly automated vehicles.
\\
\\
Complementary work \cite{Fritzsch} addresses the same project but focusses not on the safety perspective and ISO 26262 but on the implementation of the model checking. Within \cite{Fritzsch} we describe the handling of large scale model checking problems and discuss different implementations. 
\\
\\
After presenting the related work in section \ref{sec:Related Work}, we introduce a fail-operational driving system including the arbitration logic in  \ref{sec:Fail-Operation Driving Systems}. The verification process is explained in two parts in section  \ref{sec:ArbitrationLogic}. Section \ref{subsec:ModelSpecifications} describes the model; section  \ref{subsec:Specifications}. describes the formal specifications. The scope is defined by the requirements in ISO 26262. Section \ref{subsec:ScopeImplMC} further presents the implementation process. The validation approach and tool qualification are provided in section \ref{subsec:Validation} to comply with ISO 26262. The results and application are discussed in section \ref{subsec: Results and Comparison to Conventional Analysis}, including threats to validity.

\section{Related Work}
\label{sec:Related Work}
\IEEEPARstart{P}{revious} work that addresses the verification of fail-operational automotive systems exists. Additionally, some studies address the verification of automotive system by formal methods. This section reviews the literature as it relates to the compliance of the industry standard ISO 26262.
\subsection{Safety Analysis of Fail-Operational Automotive Systems}
Since safety is a major aspect of fail-operational systems, it is discussed in most related publications. However, previous research mainly focuses on reliability factors.
\\
The most comprehensive work is presented by Schnellbach \cite{Schnellbach}. This work addresses the limitations of the first version of the industry standard ISO 26262, discusses relevant aspects of fail-operational systems, such as the definition of the emergency operation time, and provides an approach to design the redundant architecture based on reliability analysis. The arbitration logic as a core element for fail-operational behaviour is addressed but not it's verification. Another comprehensive work addressing fail-operational systems is published by Sari \cite{Sari}. An architecture model is designed based on a functional safety concept. The necessity of an arbitration logic is stated but the analysis is not detailed. However, the author focuses on the analysis of dependencies in redundant elements, another highly relevant aspect of fail-operational systems.
\\
The functional verification of a fail-operational system using a formal approach is presented by Koelbl and Leue \cite{Koelbl}. The system is modeled as a state diagram at the vehicle level and takes into account an emergency operation mode after a failure. The state model is checked using statistic model checking to analyze the reliability of the concept. Although this model aggregates component states to a system-level state, it does not reflect the complexity of modern driving systems.
\\
Comprehensive work regarding the functional safety of fail-operational systems focusses on the analysis at the vehicle level, including safety, reliability, and dependability. The verification of the fail-operational behavior is addressed at the vehicle level and relevant analyses at the system level are identified. The purpose of this study is to close the gap and provide a verification procedure on system level.
\subsection{Formal Verification of Automotive Systems}
Formal verification and in particular, model checking have generated great interest in research. The research focuses on modeling, algorithms, including the handling of the state-space problem, and application. Clarke \cite{Clarke} gives an overview of the state of the art model checking. According to Singh et al. \cite{Singh}, formal methods provide accuracy, consistency and unambiguous specifications, using mathematical theorems. Thus, the application to safety critical systems has been addressed in the literature. We present publications to specifically address the safety of automotive systems. 
\\
\\
Formal verification in compliance with various aspects of the industry standard ISO 26262 has been covered in several publications. 
Leitner-Fischer and Leue  \cite{LeitnerFischer} discuss the recommendations in ISO 26262 regarding the verification and suitability of formal methods. The authors conclude that formal methods support the safety life cycle described in ISO 26262 by formalization and the proof of accuracy. In particular, formal methods provide a systematic approach and allow the traceability of safety requirements. In addition, the authors emphasize the necessity for a tool qualification. Bahig and El-Kadi \cite{Bahig} come to similar conclusions and present a general process based on formal modeling and model checking.
\\
\\
Further work addresses the application of formal methods to automotive problems and partly the industry standard ISO 26262.
\\
Abdulkhaleq and Wagner \cite{Abdulkhaleq} combine system theoretic process analysis with formal verification. Safety requirements are identified via system theoretic process analysis and formally specified. Then, model checking is used to verify a process model regarding the specifications. The basic concept is comparable to our work. It shows the benefits of a structured approach and provides a complete verification by focusing on system-level behaviour. Their approach is applied to a cruise control system with limited states and complies with the software safety requirements in ISO 26262 through traceable and formal specifications.
\\
Nyberg et al. \cite{Nyberg} present their experience using formal verification techniques including model checking at Scania, a Scandinavian truck company. The results show the applicability of model checking in general, and the challenges of the formalization of models and requirements in particular. They conclude that the verification method needs to be chosen based on system and modeling complexity. Large systems require more complexity and effort and may not be verified comprehensively. ISO 26262 is only referenced as required for a verification procedure.
\\
Another application is presented by Nellen et al. \cite{Nellen}. The authors verify a controller for a parking function using Mathworks, Simulink\footnote{\url{https://mathworks.com/products/simulink.html}}. They recognize a dependency of the computing time to the model size and the time interval to be analyzed. They conclude that model checking supports safety analysis and increases the quality of the safety requirements by using the process of formal specification. However, computing time and complexity limited the range of applications.
\\
Todorov et al. \cite{Todorov} use model checking to verify a cruise control algorithm. Once the model is formalized and implemented, verification is very efficient for large models. However, similar to Nellen et al. \cite{Nellen}, long time spans of constraints limited the analysis due to runtime problems despite the limited model size. 
\\
Other work stated model checking can be used as a complement to other verification techniques limit the time and effort. 
Da Silva et al. \cite{daSilva} combined model checking with testing and simulation. They state that formal verification is time consuming and, therefore, cannot be used extensively. Similarly, Aniculaesei \cite{Aniculaesei} uses other examples of model checking to identify test cases.
\\
Work from other industries, such as aviation, focuses on code level evaluation and rarely considers distributed functions \cite{Moy}. Thus, it is not relevant to our study.
\\
\\
In conclusion, the functional verification of fail-operational functions has not been sufficiently investigated. Related work demonstrates the applicability of model checking to ISO 26262 in the automotive industry. In particular, they examine the formal specification of requirements. However, these studies are limited and the verification of large, distributed systems has not yet been demonstrated. In addition, a complete coverage of ISO 26262 has only been addressed by \cite{LeitnerFischer} without an application. Current verification approaches cover only some factors, the need for verification and the identification of safety requirements. They do not include validation or tool certification. The objective of this study is to close this gap.

\section{Fail-Operation Driving Systems}
\label{sec:Fail-Operation Driving Systems}
\IEEEPARstart{F}{ail-operation} behaviour in an automotive context is defined as the ability of a system to be operable in the presence of a failure. Such behaviour is required when it is not immediately possible to reach a safe state by deactivating the system \cite{Isermann}. That corresponds to fault-tolerance according to the industry standard ISO 26262 \cite{ISO26262}. Since safety is a key requirement, the fall-back operation time and functionality are limited in response to a system failure \cite{Schnellbach}. 

\subsection{Fail-Operation Safety Goals}
Fail-operational behaviour requires redundant architectures because a fall-back level is necessary in case of a failure resume the driving task \cite{Niedballa}. The necessity of such fail-operational system behaviour is defined as part of the safety concept, resulting from the safety goals. Safety goals are safety requirements at the system level and are identified from a hazard and risk analysis. These safety goals are given in our analysis and included for comprehensibility. Table \ref{tab:Safety Goals} shows the applicable safety goals specified according to ISO 26262 \cite{ISO26262} and the literature \cite{Stolte} \cite{Koelbl}. Those safety goals include the activation, deactivation and operation of the highly automated driving mode. The fail-operational behaviour is described in case of failure. Each safety goal includes the relevant specification, a respective integrity in accordance with ISO 26262 and the fault-tolerant-time interval (FTTI). The fault-tolerant time interval describes the time span after which the failure becomes critical and a transition to a safe state must have been achieved. The safe state can either be a state without any hazard or a state where the systems’ integrity corresponds to hazards.
\begin{table}[!t]
\caption{Safety Goals for a Fail-Operational driving System}
\label{tab:Safety Goals}
\setlength{\tabcolsep}{0.5em}
\begin{tabular}{p{0.04\textwidth}|p{0.27\textwidth}|p{0.06\textwidth}|p{0.05\textwidth}}
\hline
& &\\[-7pt] 
\textbf{ID} & \textbf{Safety Goal} & \textbf{Integrity} & \textbf{FTTI} \\
\hline
& &\\[-5pt]  
SG 1 & The function must not be activated when the components do not signal readiness. & ASIL B & 0 ms \\[15pt] 
SG 2 & The function must not be deactivated falsely. & ASIL D & 0 ms \\[15pt]  
SG 3 & A collision by leaving the trajectory in the nominal operation must be prevented. & ASIL D & 0 ms \\[15pt]  
SG 4 & The arbitration logic must activate a functioning fall-back operation in case of a failure. & ASIL D &  200 ms \\[15pt]  
SG 5 & The system must decelerate after a switch-over. & ASIL B &  200 ms \\[15pt]  
SG 6 & A collision by leaving the trajectory in the fall-back operation must be prevented. & ASIL B & 0 ms \\
\end{tabular} 
\end{table}
\\
Activation is only allowed if no failure is present and no false deactivation is possible. Furthermore, the driving task must be performed in accordance with the respective integrity. Due to the possible severity, that is ASIL D during normal operation. As specified by ISO 26262, the risk is limited by deceleration. That is stated in Table \ref{tab:Safety Goals} safety goal number 5 (SG 5). Therefore, the fall-back operation meets the specifications of ASIL B. The transition between the nominal- and fall-back operation is ensured by an arbitration logic, which must determine the functional channel correctly with the specifications of ASIL D and within the fault-tolerant time interval. That means systematic failures in the logic must be prevented and all sequences of a switch-over must be within the time interval. 
\subsection{Fail-Operational Architecture}
In this section we present the architecture of a fail-operational driving system in accordance with the safety goals shown in Table 1. Electric and electronic (E/E) architectures at the vehicle level for fault tolerant driving systems have been discussed in the literature by Kron et al. \cite{Kron}, Niebdalla and Reuss \cite{Niedballa}, Schnellbach \cite{Schnellbach} and Sari \cite{Sari}. All of these E/E architectures use dynamic redundancy but differ in their fall-back configurations. Dynamic redundancy, in contrast to static redundancy, uses fewer components and a self-diagnostic system for each channel \cite{Sari} \cite{Isermann}. 
\\
\\
This study, in collaboration with the BMW Group, investigates the fail-operational driving system of BMW as described by Kron et al. \cite{Kron} The system is already in road testing and therefore in a mature stage of development. The driving system consists of two redundant channels, each capable of conducting the driving task. Figure 2 shows the architecture.
\begin{figure*}[t!]
\centering
\includegraphics[trim={0mm 0mm 0mm 0mm},clip,scale=1.7]{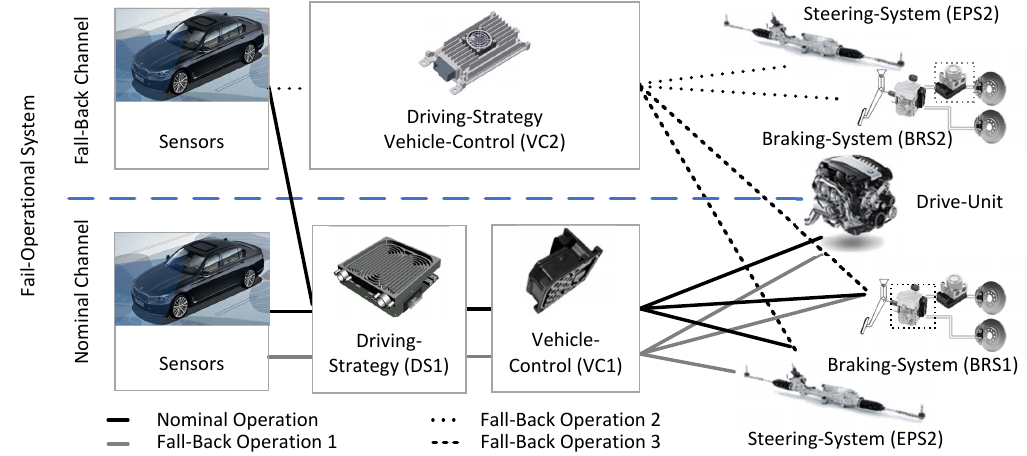} 
\caption{Architecture of a Fail-Operational Driving System \cite{Kron}}
\label{fig:failOpArchitecture}
\end{figure*}

\begin{figure}[!t]
\centering
\includegraphics[trim={0mm 0mm 0mm 0mm},clip,width=3.5in]{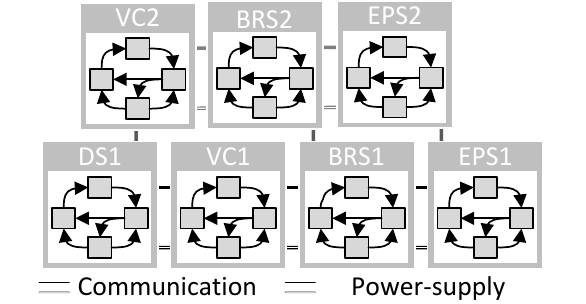}
\caption{Arbitration Logic as Composite of State-Machines}
\label{subfig:SMCpompund}
\end{figure}

The nominal channel is in the bottom and consists of sensors and modules to determine the driving-strategy (DS1), the vehicle-control (VC1) and actuators, each partitioned on separate electronic control units (ECUs). The fall-back channel at the top includes sensors, a single ECU for the driving strategy, and the vehicle control (DS2). The braking-systems (BRS) and electronic power steering-system (EPS) are redundant in each channel, whereas the drive-unit is only integrated into the nominal channel. That is possible since fail-silent behaviour is safe in the drive-unit and therefore fault-tolerance is not required. The system is able to conduct the driving-task in nominal operation mode and three fall-back operation modes. These define the system configurations. In case of a failure, a backup operation is activated. The current fall-back operation mode depends on the failure. The nominal operation and the fall-back operation mode 1 is conducted on the nominal channel in case of a failure in the fall-back channel. Fall-back operation mode 3 is the operation on the fall-back channel, in case of a failure in the nominal channel. Since the primary actuators comprise a greater range of functions, a prioritization system is implemented \cite{Kuemmel}. That means the actuators are used in the corresponding channel, and in addition, the primary actuators (BR1, EPS1) can be controlled by the control systems on the fall-back channel (VC2) which is shown in operation mode 2. A direct loss results from an undetected failure in the nominal operation mode or a second failure in the fall-back operation. The switch-over is ensured by an arbitration logic that is distributed throughout the cooperative modules in the applicable control units. In addition, the driver is requested to take over in case of a failure. If no take over is conducted after a certain time, the vehicle executes an emergency operation, such as a deceleration.
\\
\\ 
The arbitration logic is implemented by a set of distributed state machines, shown in Figure \ref{subfig:SMCpompund}. The naming corresponds to figure \ref{fig:failOpArchitecture}. The logic has to ensure the fail-operational behaviour and thus determines the operation mode, i.e. nominal or fall-back. The composite of the state machines conducts the switch-over cooperatively. Each state machine is partitioned on an electronic control unit (ECU) and is connected to at least one communication bus and power supply.

\begin{figure}[!b]
\centering
\includegraphics[trim={0mm 0mm 0mm 0mm},clip,width=2.0
in]{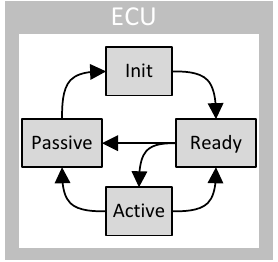}
\caption{Exemplary State-Machine}
\label{subfig:SMsingle} 
\end{figure}

Figure \ref{subfig:SMsingle} displays a single state machine consisting of an initial state (\textit{Init}) and states representing \textit{ready}, \textit{active} and \textit{passive} operations. The state machines are connected via the transition conditions.

After the start-up of the control-unit, the state machines are in in the initial state (\textit{Init}). When the diagnostics are completed and no failure has been detected, the state-machines switch to \textit{Ready}. By activation of the highly automated driving mode, the state-machines on the nominal channel switch to \textit{Active}, whereas the state-machines at the fall-back level stay in \textit{Ready} state. In case of a failure, the respective state-machines switch to \textit{Passive} and the others react by deactivating the failed channel and activating the respective fall-back channel. That also occurs if the other state machines do not receive any signal from the failed state machine. As mentioned, a prioritization is considered, which enables the operation of a primary system in combination with the fall-back channel.
\\
\\
It is assumed that the fail-operational behaviour is independent of the failure leading to the fall-back operation and the operation mode. In the context of ISO 26262, that means dependent failures must be prevented.  

\section{Verification of the Arbitration Logic}
\label{sec:ArbitrationLogic}
\IEEEPARstart{I}{n} this section, we present the verification of the arbitration logic. We first explain the overall requirements of ISO 26262 and then discuss the individual sections in detail. 
\\
The development of critical safety software with regard to industry standards provides comprehensive and trackable means to fulfill the safety goals. Thus, a structured evidence in form of a verification is provided. The level of detail is defined according to industry standard using integrity and failure probabilities. Therefore, the method must be complete and well-structured with regard to failures and its relevant behaviour. The safety argumentation procedure must be consistent within the subsystems. A validation must be provided and when tools are used, a tool qualification needs to be conducted as well. The procedure and implementation must be independently reviewed to assure compliance with ISO 26262.
\\
\\
We used the established and well documented open-source tool NuSMV\footnote{\url{http://nusmv.fbk.eu/}} to check the model. This tool can be applied to bounded and unbounded models and formal specifications using linear-time and computational-tree logic. The algorithm is based on the construction of a behaviour-driven-diagram and reachability algorithms \cite{Cavada}. The optimisation strategy of other tools with regard to computing time is to detect failures earlier. Since our goal is complete verification, this strategy is not applicable to our research. Instead, we implement the appropriate state-oriented syntax. See documentation \cite{Cavada} \cite{CavadaB} for a detailed explanation.
\\
\\
In this section, we first construct the model. Then, we explain how to check the model by formulating formal specifications and explaining the implementation. Additionally, we present our validation procedure and explain the tool qualification process to meet the requirements of ISO 26262. In the last section, we present the verification result of a fail-operational driving system at BMW and discuss its application including a comparison to other safety analyses.

\subsection{Modelling of the arbitration logic}
\label{subsec:ModelSpecifications}
This section discusses the modeling of the fail-operational system presented in section \ref{sec:Fail-Operation Driving Systems}. The modeling process of the system behaviour is also affected by the definition of relevant failures. The relevant failures of the architecture and external signals need to be considered in the implementation of the state machine logic. We conduct the failure identification by an exploratory approach following a failure mode and effect analysis and consider consistency of the related safety analysis for a comprehensive safety argumentation.
\\
\\
Internal as well as external failures which trigger a reaction in the arbitration logic are considered in the model. Internal failures include architectural failures, such as control units, power supply, and communication. External failures are  signals which are not builded in the arbitration logic. These are reported functional failures and signals for activation and deactivation. Safety analysis at the software level verifies the functional trigger. Safety analysis at the component level\footnote{Component level is equivalent to ECU level in our context} ensures a shut-down of the ECU, and the separation of the voltage source for safety- critical failures. That includes lower-level software failures, hardware failures, CPU parts failures and communication failures. Central Processing Unit failures and power supply failures are equivalent to a loss of communication. Both, power supply and communication failures can affect either single connections or the complete wiring for both sender and receiver. Sender and receiver failures are included in ECU failures. The corruption of signals is inhibited with integrity. That implies end-to-end protection and leads to identical failure modes resulting in no transmission.
\\
\\ 
The purpose of the arbitration logic is to ensure fail-operational behaviour and to determine the operation mode with regard to the system configurations. As explained, this includes the activation, deactivation, and transition to a fall-back operation in case of a failure. The model consists of modules for each state machine and communication connection along with a main module, where the instantiation occurs.
\\
\renewcommand{\lstlistingname}{Implementation}
\begin{lstlisting}[caption={},label={Listing:StateMachine}, float=hbt, caption={Implementation of a state machine by means of the driving strategy}]  
MODULE M_DS1 (VC1, VC2, EPS1, BRS1, Failure$_{Function DS1}$, Activation)
 VAR
  DS1: {Init, Ready, Failure, Active};		
 ASSIGN		
  init(DS1) := Init;		
  next(DS1:= 
   case			
    DS1 = Init & !Failure$_{Function DS1}$: {Ready};
	DS1 = Ready & (Activation & (VC1&DS2) = Ready): {Active};
	DS1 = Ready & Failure$_{Function DS1}$: {Failure};
	DS1 = Active & !Activation: {Ready};
	DS1 = Active & ((Failure$_{Function DS1}$|VC1|BRS1|EPS1)=Failure): {Passive};
	DS1 = Failure & !Failure$_{Function DS1}$: {Init}
   TRUE: DS1;
   esac;		
\end{lstlisting}
Each state machine is modeled as an instance with corresponding inputs as shown in driving strategy 1 \textit{DS1} in implementation \ref{Listing:StateMachine}.

The code in implementation \ref{Listing:StateMachine} displays a simplified state machine for the driving strategy similar to Figure \ref{subfig:SMsingle} and uses a switch case statement. The state machines read the other states, symbolised by the respective variables VC1, VC2, and so on, switch accordingly. After the system has been started and no failures have been diagnosed, the state machine signals \textit{Ready}. 
The driving strategy switches to \textit{Active} when an activation is triggered and the other state machines signal readiness. If a failure occurs, the system transitions to the \textit{Passive} state. In the \textit{Active} state, the states of the other state machines on the channel are additionally evaluated to ensure a deactivation of the complete functional channel. If the failure is cured, the state machine switches to \textit{Init} but does not directly reactivate. Additionally, the driver may deactivate the operation in any state, even if it is not displayed. In the corresponding implementation \ref{Listing:StateMachine}, system specifications are commented in the code to support comprehensibility. 
\\
\\
Furthermore, the system architecture must be evaluated because the interface to the hardware affects functionality. As previously stated, that includes the partitioning on CPUs, power supply, and communication as shown in Figure \ref{subfig:SMCpompund}. Multiple buses are used to achieve communication between the state machines. Each communication link between the dispatch of a state to an individual state-machine is modeled. For robustness, the debouncing of signals is used in the system. If a signal is not received, the receiver assumes the signal is faulty after a signal has not been detected multiple times. Since the signals are discrete and end-to-end protection with adequate integrity is implemented, the failure mode is evaluated as no signal transmission. The power supply can affect a complete power circle or single supply module and lead to a shut-down of the corresponding CPUs, also resulting in no signal transmission. Therefore, we model architectural failures as a trigger for communication failures. Implementation \ref{QCode:Kommunikation} shows an example of a communication failure of the connection between a state machines of the vehicle-control unit (VC1) and the driving-strategy (DS1) as a result of a failure in the power supply.

\begin{lstlisting}[caption={Implementation of communication as architectural failure},label={QCode:Kommunikation}, float=hbt]  
init(Comm$_{DS1-VC1}$) := 	Init;
 next(Comm$_{DS1-VC1}$) := 	
 case 				
  Failure$_{Energy}$ & t$_{debounce}$ >= 3 : {Failure};
  Failure$_{Energy}$ & t$_{debounce}$ < 3 : {Comm$_{DS1-VC1}$};	
 TRUE: Comm$_{DS1-VC1}$; 		
 esac;	
\end{lstlisting}
The communication signal from the vehicle-control unit (VC1) to the driving-strategy (DS1) bypasses the state of VC1 if no failure has been diagnosed or if the debounce-time has not yet been reached. If that is not the case, the corresponding state machine switches to the \textit{Passive} mode. Similarly, failures in control units, communication buses, and communication links are evaluated in the same fashion. All communication connections are combined in one communication bus.
\\
\\
The initiation of the state machines and the communication bus occurs in the main module. Implementation \ref{Listing:Main} shows the structure.
\begin{lstlisting}[caption={Main Module},label={Listing:Main}, float=hbt]  
Failure$_{Function DS1}$
...
Bus: M_Bus(DS1,VC1,BR1,EPS1,VC2,BR2,EPS2)
DS1: M_DS1 (Bus.VC1, Bus.VC2, Bus.EPS1, Bus.BRS1, Failure$_{Function DS1}$, Activation)
\end{lstlisting}
In the first part, all variables and failures are initiated. The second part shows the initiation of the bus, which first includes the communication connection. In the second part, the state machines are shown. The state of each machine is passed via the bus, instead of directly showing its state. That enables the failure effects to be modeled. 
\\
\\
The model depicts a simplified representation of the arbitration logic. First, we model the communication connections independently of the actual bus system. In particular, asynchronous communication behaviour is not modeled. Furthermore, the computation is conducted cyclically and synchronously, which generally does not represent the behaviour distributed systems. A delay of one time-step is produced when the communication implemented via the bus module. We explain the significance of these results in Section \ref{subsec: Results and Comparison to Conventional Analysis}.
\subsection{Formal Specification of the Safety Goals and Requirements}
\label{subsec:Specifications}
In addition to modeling, system constraints need to be formally specified to verify the system. Safety requirements for the arbitration logic are formulated in a similar fashion to Abdulkhaleq and Wagner \cite{Abdulkhaleq}. The requirements are derived from the safety goals in Table \ref{tab:Safety Goals}. Then, they are translated from a narrative form to a formal specification which can then be used for the model checking procedure.
\\
\\
The requirements describe how the system reacts to an external trigger or a failure, the operational modes, by the transitions of each state machine. Specification \ref{QCode:RB1} shows the definition of the fall-back mode 1 (FB1).
\renewcommand{\lstlistingname}{Specification}
\setcounter{lstlisting}{0}
\begin{lstlisting}[label={QCode:RB1}, caption={Definition of fall-back operation mode},float=hbt]  
init(FB1) := FALSE;
 next(FB1) :=
 case
  (DS1 & VC1 & EPS1) = Active &
  (VC2 & BR2 & EPS2) != Active &
  !(VC2 & BR2 & EPS2 = Ready) : True
  esac;
\end{lstlisting}

The state machines in the nominal channel are all in \textit{Active} state; the ones in the fall-back channel are not in \textit{Active}. At least one state must be unequal to \textit{Ready} since only a detected failure in the fall-back channel leads to the fall-back operation.
\\
\\
Important safety requirements include activation and deactivation, as well as the actual fail-operational behavior to react to failures. The following section describes the formal specifications of the requirements defined from the safety goals.
\\
\\
Activation is only possible when all state machines signal readiness and a request for activation has been sent. The requirements correspond to Safety goal No. 1, which is ready to be implemented. Therefore, no further level of detail is necessary. 
\begin{lstlisting}[caption={Condition for Activation}, label={SpecActiv},float=hbt]
G (NO -> (O ((DS1 & VC1 & ... & EPS2) 
   = Ready & Activation = 1)))
\end{lstlisting}

The formal specification states that, once the normal operation is activated all state machines are in \textit{Ready} state and the trigger for the \textit{Activation} was present.
\\
\\
The failure-tolerant behaviour reacts to a failure and is thus inevitable for fail-operational systems. The safety goals (e.g. Table \ref{tab:Safety Goals} Safety Goal 4) do not specify the explicit configuration of the system dependent of the failure. However, that is necessary for verification since the fall-back operation modes include degraded functionality and thus the operation is prioritized. In addition, we specify the requirements for single and dual-point failures, since the target operation mode might differ. The justification for this is provided later. 
\\
As mentioned before, we differentiate between failures that are followed by a direct reaction, and failures that are debounced for robustness. The requirement states that for specific failures in the nominal channel, the system must switch to the fall-back channel.
\begin{lstlisting}[caption={Switch-over to fall-back after single point failure}, label=SFRB2,float=hbt]
G (((DS1|VC1) = Failure | 
  Failure$_{NC}$.t$_{debounce}$=3)
  -> (G [FTTI-5,FTTI+5] (FB2))) 
\end{lstlisting} 
Thus, the formal specifications, include the state machines in the nominal channel which lead to fall-back operation 2 and the corresponding debounced failures which occurred in the supply, etc. In this case the arbitration logic must activate the fall-back operation using the fall-back channel. That has to occur within the fault-tolerant time interval for all cases (G, globally).
\\
For double-failures, we need a condition to verify that more than one failure has occurred (see Specification \ref{DFRB2}). The upper limit is set by the failure combinations, a subset of failures, which are verified via an negated exclusive or construction. A single point failure needs to be excluded. 
\begin{lstlisting}[caption={Switch-over to fall-back after multi-point failure}, label=DFRB2,float=hbt]
G ((((DS1|VC1|) = Fehler 
  | Failure$_{NC}$.t$_{debounce}$=3)
  & !(DS1 xor VC1) = Failure xor Failure$_{NC}$.t$_{debounce}$=3)) 
  -> (G [FTTI-5,FTTI+5] (FB2))) 
\end{lstlisting} 
Other specifications are not directly identified from the safety goals but result from the safety concept of the arbitration logic. 
It is also necessary to prevent the system from toggling between the channels and operation modes even in the case of oscillation failures to ensure a stationary state.
\begin{lstlisting}[caption={Prevention of Toggling and Reactivation},float=hbt]
G ((FB1 -> !FB2) & (FB2 -> !FB1) &
  ((FB1 | FB2) -> !NO)) 
\end{lstlisting}
Therefore, we specify that a switch from the primary fall-back operation to the secondary fall-back operation and vice versa as well as a switch from any fall-back operation back to the nominal operation mode are both defined as failures.
\\
We also exclude an operation in more than one operation mode. This can also be covered in the definition, such as Specification \ref{QCode:RB1}. 
\begin{lstlisting}[label=OneMode, caption={Exclusiveness of operation modes},float=hbt]
G !(NB & FB1) & !(NB & FB2) & !(FB1 & FB1)))
\end{lstlisting}
Specification \ref{OneMode} must be valid globally and ensures that only one mode is active.
\\
\\
Deactivation, corresponding to Safety goal No. 2, is formulated similarly to the Activation in Specification \ref{SpecActiv}. The target state for each state machine is not equal to active when a corresponding signal is received.  
\subsection{Extent and Implementation of the Model Checking}
\label{subsec:ScopeImplMC}
The scope of analysis is given by the requirements of ISO 26262 \cite{ISO26262}. First, we define the scope and extent of the analysis. Then, we explain the implementation and in particular, the assignment of specifications to failures.
\\
\\
In ISO 26262, fail-operational fault tolerance is defined as the ability for a functionality to operate in the presence of one or more faults. That means the analysis needs to cover at least single point failures. The ISO 26262 considers multi-point failures with higher order than two as safe, unless the safety concept requires the contrary. In dual-point failures, plausibility has to be evaluated based on the probability of occurrence and dependencies. Since a fail-operational driving system fulfills both requirements, double-faults do not need to be fully controlled. However, the specifications should cover all reactions and the analysis of dependent failures is based on the failures leading to a loss of functionality. All double-faults scenarios are evaluated.
\\
\\
Not every constraint is relevant to every failure combination. To limit the constraints to the corresponding subset of each failure combination, we separate the failure scenarios. In accordance with ISO 26262, failure combinations are limited to second order to further counteract the state space explosion. In addition, this separation allows us to verify the target channel without limiting the failure scenarios in the constraints. Table \ref{tab:Failurematrix} shows a subset of the matrix to determine the relevant constraints for each failure combination, depending on the target condition.
\renewcommand{\arraystretch}{1.4}
\begin{table}[]
\centering
\caption{Matrix of failure combinations and target channel}
\label{tab:Failurematrix}

\begin{tabular}{lccll}
\cline{1-3}
\multicolumn{1}{l|}{\diagbox{\textbf{1st Fault}}{\textbf{2nd Fault}}}                                                          & \multicolumn{1}{c|}{\begin{tabular}[c]{@{}c@{}}Steering \\ Function 1\end{tabular}}       & \begin{tabular}[c]{@{}c@{}}Power Supply\\ fall-back\end{tabular}      &  &  \\ \cline{1-3}
\multicolumn{1}{c|}{\begin{tabular}[c]{@{}c@{}}Steering\\ Function 1\end{tabular}}   & \multicolumn{1}{c|}{\begin{tabular}[c]{@{}c@{}}fall-back operation \\ mode 3\end{tabular}} & Inactive                                                             &  &  \\ \cline{1-3}
\multicolumn{1}{c|}{\begin{tabular}[c]{@{}c@{}}Power Supply\\ fall-back\end{tabular}} & \multicolumn{1}{c|}{Inactive}                                                             & \begin{tabular}[c]{@{}c@{}}fall-back operation \\ mode 1\end{tabular} &  &  \\ \cline{1-3}
                                                                               & \multicolumn{1}{l}{}                                                                      & \multicolumn{1}{l}{}                                                 &  & 
\end{tabular}
\end{table}
The primary faults are listed vertically; the secondary faults are listed horizontally. Single-point failures are listed on the diagonal line. The other cells list double point failures. Table \ref{tab:Failurematrix} shows that a failure of the primary steering function leads to a fall-back backup operation in fall-back mode 3. This leads to specification \ref{DFRB2}. The matrix is not symmetric in general. For example, an emergency braking function might only be triggered by a certain order of events, or by a prioritization of components. 
\\
\\
The failure times for the primary and secondary failure are limited in order to verify the order of the failures. The failure tolerant time interval refers to the last possible occurrence of the second failure and is implemented by a counter. As shown in Section \ref{subsec:ModelSpecifications}, we verify the state for a time interval to ensure a stationary state. Therefore, we can use bounded model checking which verifies the termination of the model checking. The search-depth is defined according to the stationary state. All failures can occur at any time for any possible time span during the defined interval and do not disappear. This results in an approximated asynchronous behaviour, which egalises the simplifications of the model.
\subsection{Validation and Tool Qualification}
\label{subsec:Validation}
Resilient implementation affords a validation to ensure correctness. Furthermore, ISO 26262 \cite{ISO26262} requires a tool qualification as proof of reliable verification results.

\subsubsection*{Validation Procedure}
The validation procedure needs to address the model and the formal specifications.
\\
\\
We use individual failure combinations as stimuli and verify the model behaviour. The target state are evaluated for every failure combination according to Table \ref{tab:Failurematrix}. However, complete model validation is conducted during verification. For that purpose, a precondition is that constraints are used which include any failure, debounced in the relevant cases. Those constraints are similar to the specifications shown in Specification \ref{SFRB2} and \ref{DFRB2}. That only holds when violations are detected correctly, as addressed in the following section.
\\
The validation of the formal specifications requires more manual effort. We use failure injections based on equivalence groups in accordance with Hoffmann \cite{Hoffmann} to show the appropriate formulation. In addition, expert reviews ensure completeness. The formal specifications shown in Subsection \ref{subsec:ModelSpecifications} consist of preconditions and target criteria; we validate both to identify failures of the arbitration logic. First, we identify a subset of failure combinations which cover all specifications. Then, we conduct the validation steps listed in Table \ref{tab:Validationmatrix}.
\renewcommand{\arraystretch}{1.4}
\begin{table}[t]
\centering
\caption{Matrix for validation of Specifications}
\label{tab:Validationmatrix}
\begin{tabular}{lccll}
\cline{1-3}
\multicolumn{1}{l|}{\diagbox{\textbf{condition}}{\textbf{target criteria}}}                                                              & \multicolumn{1}{c|}{true} & \multicolumn{1}{c}{false}                                                                     &  &  \\ \cline{1-3}
\multicolumn{1}{c|}{true}  & \multicolumn{1}{c|}{ok} & \multicolumn{1}{c}{!target criteria}                                                          &  &  \\ \cline{1-3}
\multicolumn{1}{c|}{false} & \multicolumn{1}{c|}{-}    & \multicolumn{1}{c}{\begin{tabular}[c]{@{}c@{}}!target criteria \\ \& !condition\end{tabular}} &  &  \\ \cline{1-3}
                            & \multicolumn{1}{l}{}      & \multicolumn{1}{l}{}                                                                           &  & 
\end{tabular}
\end{table}
As a baseline situation, we use the specifications without any failure injections. The verification result should be \textit{no violation}. There is no benefit in solely manipulating the condition since we check for violations. Not fulfilling the condition would not change the result. However, a negation of the target criteria would cause a failure. Therefore, we can show that it is formulated correctly. We negate both the target criteria and the condition to ensure the accuracy of the condition. This prevents a failure for the same failure combination since the condition is not fulfilled. We conduct reviews to ensure the completeness of the conditions. This allow us to validate only a subset of conditions.
\subsubsection*{Tool Qualification}
\begin{figure}[!b]
\centering
\includegraphics[trim={0mm 0mm 0mm 0mm},clip,width=3.0in]{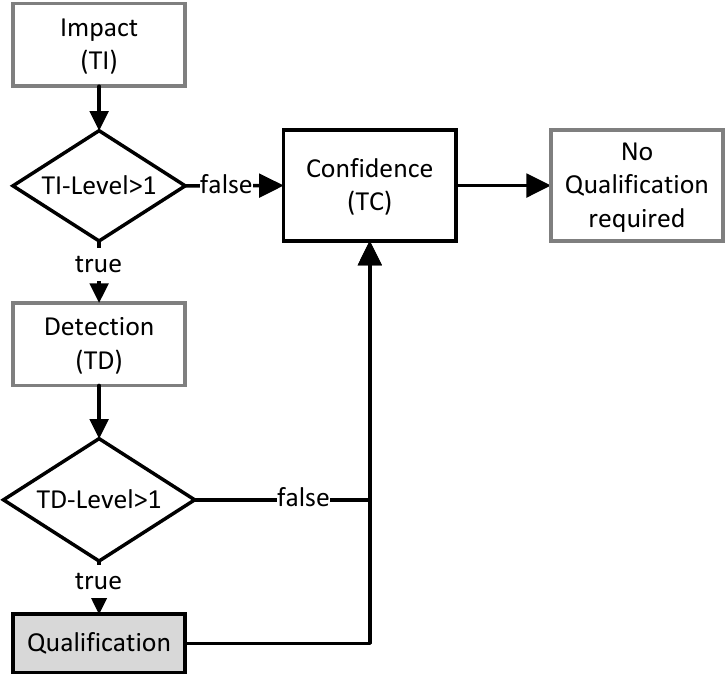}
\caption{Evaluation of the Tool Qualification Level \cite{ISO26262}}
\label{fig:ToolQual}
\end{figure}

The ISO 26262 requires a qualification of the software tools used for the development of safety critical products. The qualification is necessary whenever a work product of the safety life-cycle, such as a verification, relies on software tools. That is the case in our study. It needs to ensure the resilience of the results. The qualification needs to incorporate NuSMV as well as the control program to depict the relevant checking cases. We first explain the qualification procedure according to ISO 2626 and then how to implement it in our use case.
\\
\\
The ISO 26262 provides a guideline for the qualification process, which covers an analysis and verification of the impact, either through a validation procedure or a guideline for the development.
\\
At first, the relevant use cases and the respective failure cases are identified. Thus, the software-tool is treated as a black box system. An evaluation of the tool impact (TIL) and the tool detection level (TDL) resulting in a confidence level (TCL) is used to determine the necessary qualification measures. Figure \ref{fig:ToolQual} gives an overview of the evaluation process as described in ISO 26262.  
 
The levels consist of several gradations. However, only level 1 is adequate for safety critical utilization. At first, the tool impact level is determined for every use-case. If there is a safety critical impact because of failures during the verification process, the tool detection level is evaluated. If safety critical failures are not systematically and comprehensively detected, qualification is afforded. This can be granted by developing the tool in compliance with ISO 26262 either by depending on the complying integrity or validation procedure, evaluating the development process validation or having confidence by its usage.
\\
\\
We illustrate the procedure for our verification tool in the following section. The relevant use case is the detection of violations. Detecting a false-negative is critical because it would lead to potential faults in the product. Detecting a false-positive requires more effort in the evaluation but it is not a safety critical issue. Reasons for a false-negative verification of critical failures are because the model checking procedure did not occur or because a wrong application of the specifications was used.
\\
Since we use an available tool, we conduct a qualification by validation using a test-set. The aim of the tool qualification is the proof of the tool accuracy and not the full coverage of all test cases; a subset is sufficient. The first relevant cause, no conduction of the model checking, is evaluated by defining the target state in the formal specifications and the stated validation of the specifications. Again, this is only applicable under the premise that failures are detected. Therefore, we want to prove that this is the case during the verification via model checking. We use failure injection in the state-machines, implementation of the communication connections, and manipulation of the matrix for failure combinations. In addition, faulty usage of the tool is minimized by restricting the circle of users. 
\\
Compliance with ISO 26262 is obtained by validation and tool certification. Reviews confirm this process. Our tool can thus be utilized for the verification of the safety critical system in accordance with the industry standards.

\subsection{Results of the model checking}
\label{subsec: Results and Comparison to Conventional Analysis}
In the following section, we present the results and findings of the verification process. We applied our procedure to a fail-operational driving system of BMW. The architecture is similar to that presented in Section \ref{sec:ArbitrationLogic}. The system consists of seven coupled state machines with thirty individual states. Those result in more than six thousand potential combinations. Through prioritization, four operation modes are possible (the nominal mode and three fall-back modes). In addition, five communication buses and 25 individual signal connections are evaluated. The system's failure matrix has a size of $48\times48$, including the power supply. The failure types are described in subsection \ref{subsec:ModelSpecifications}. The failure matrix is asymmetric due to the prioritization of primary actuators.
\\
First, we present the findings and results. Then, we evaluate the application of the verification approach.

\subsubsection*{Contribution to the Safety Case}
We were able to verify the system’s functional safety in accordance with the industry standard for functional safety, ISO 26262 through the verification process. The verification process directly enhances a fail-operational driving system in accordance with ISO 26262 requirements. This is a mandatory requirement for the homologation of automotive systems. Compliance with industry standards was confirmed by an independent review. Further testing verified the results.
\\ 
We improved the quality of the system specifications by unambigous requirements unambigious. However, we did identify one case, which lead to an unspecified state. This theoretical failure which requires specific timing during the recovery of a communication bus, lead to the activation of two operation modes. Other than that, no unspecified or unambious behaviour was detected. This demonstrates the high level of maturity of this project.

\subsubsection*{Application of the Approach}
\begin{figure}[t!]
\centering
\includegraphics[trim={250mm 120mm 250mm 110mm},clip,scale=0.75]{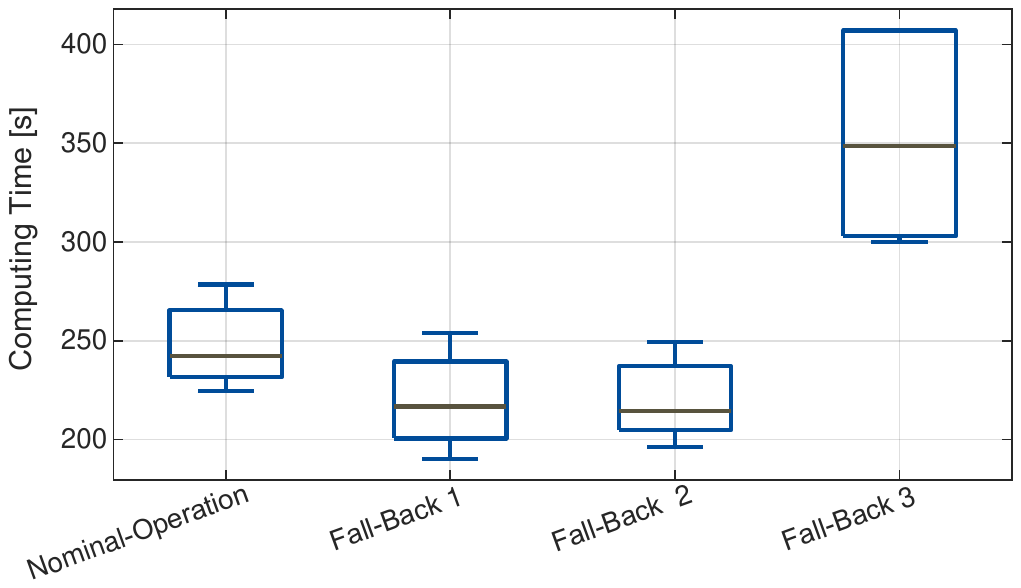} 
\caption{Computing time for the model checking of single-point failures}
\label{fig:ResEinzelfehler}
\end{figure}

\begin{figure}[t!]
\centering
\includegraphics[trim={250mm 120mm 250mm 110mm},clip,scale=0.75]{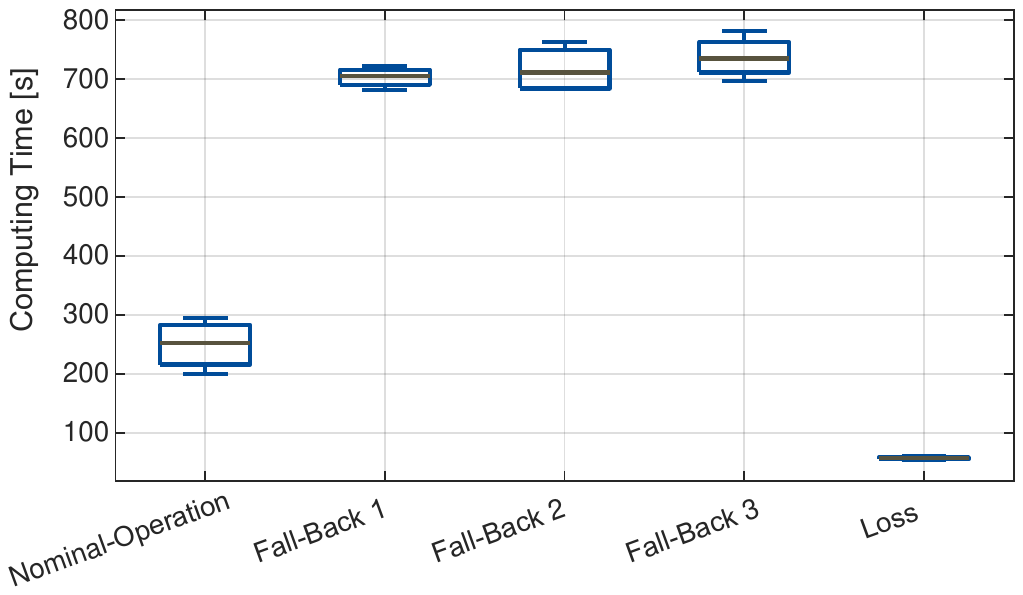} 
\caption{Computing time for the model checking of double-point failures}
\label{fig:ResDoppelfehler}
\end{figure}

In addition to improving the evaluation process of critical safety requirements, we evaluated the implementation of the verification system. The application consists of two main aspects: implementation effort and the calculation time. As stated, we used the open-source tool NuSMV \cite{Cavada} for verification and a simple control program to depict failure cases.
\\
The	implementation	of	the	model	was	fairly straightforward	since the state-based	formulation	of the	systems	requirements	is	easily transferable to	the model checking syntax. The implementation of the architecture afforded an adequate concept, but required limited effort since all architectural failures were treated similarly through a loss of communication.
\\
The identification of the formal specifications of the requirements required more effort. This is reflected in the literature \cite{Nyberg} \cite{Nellen}. Knowledge of the LTL respective CTL syntax, the model checking approach, and the system to be modeled is required. We iteratively reworked the constraints using the validation procedure and checked for plausibility in the case of deviations. Despite our understanding of the issues, identifying the formal specifications took a significant amount of time due to the system’s complexity. Modifications to the requirements or the model require rework of the specifications as well.
\\
\\
The computing time for a given model depends on the amount of constraints, their formulation, and the amount of possible states. The latter is determined by the model and the combination of failures. The formal specifications are depicted by the resulting operational mode, as described in Section \ref{subsec:ScopeImplMC}.
\\
\\
Figure \ref{fig:ResEinzelfehler} and \ref{fig:ResDoppelfehler} show the distribution of computing times in seconds for single and double failures depending on the subsequent operation mode using boxplots. The whiskers display the minimum and maximum values, whereas the box shows the respective quartiles. We conducted the verification process on a conventional computer (2.4\,GHz, 8\,GB RAM, 4\,cores, 64\,bit). We evaluated all failures and failure combinations resulting in 48 different single failures and 2256 double faults. A failure case includes the verification of all possible occurrences of that failure and, therefore, requires multiple sequences to be checked. These cases are not equally distributed among the operation modes, but at least ten single failures and 120 double failures were evaluated.
\\
Figure \ref{fig:ResEinzelfehler} displays the computing time in seconds for the single failures, depending on the operation mode as a reaction to the failure. In that case, the amount of constraints differ only slightly. The computing time for the nominal channel is slightly higher than the computing time for failures leading to fall-back 1 or fall-back 2. The mean time is between 220 s and 240 s. That increases when verifying failures leads to fall-back mode 3, which is the operation in the fall-back channel. The increased transition time is due to the amount of required state transitions, the deactivation of the nominal channel, and the activation of the fall-back channel. The variance of the computing time of failures with the same target channel depend on the failure and the resulting state sequences. For example, a loss of the power-supply affects all the connected state machines. The checking of the single-point failures took three and a half hours.
\\
The computing time is much higher for double-failures. This is due to the larger state space spanned by the failure combinations.  
Failure combinations not leading to a switch over take approximately the same computing time for single and double-failures since the state space only increases with increased failure states. In case of a switch-over the computing time is longer. The mean time increases by more than 500\,s. This results from the increase in states and sequences because the constraints are formulated similarly. In contrast, a system failure is verified under 100\,s. In that case, only one constraint, the attainment of the target state including the deactivation of all state-machines, is verified. The checking of double-point failures took seven days.
\\
\\
A comparison to other safety analysis techniques is necessary when evaluating the application of our approach. Both inductive and deductive analysis are used in the automotive industry to comply with ISO 26262. Since the goal in our use case is to prevent systematic failures and investigate failure reactions, probabilities are not relevant. It is common in the automotive industry to use an inductive method such as the failure mode and effect analysis. Although this method evaluates relevant failures and stationary states, it simply cannot cope with the amount of possible state transitions that may easily reach billions of possible sequences. Furthermore, we want to mention that such safety analysis often takes months and may need to be carried out over the complete development of the product live-cycles since they do not provide fast feedback which can be directly incorporated into the product development.
\subsubsection*{Threats to Validity}
Threats from insufficient validity may potentially lead to unreliable or faulty results which affect the internal validation. The threats to external validity are also discussed.
\\
\\
Potential failures may arise in each step of the approach, during the modeling and formal specifications procedure, as well as during the validation process. The scope of the model is based on the identification of failures in compliance with ISO 26262. We use a structured procedure to address these threats. The ISO standard requires verification and validation of the specifications as well as confirmation of every work product by independent reviews. The review includes boundary conditions and simplifications. Simplifications relate to the system behavior and in particular, the corresponding timing, as mentioned in Section \ref{subsec:ModelSpecifications}. First, the model is implemented with synchronous behavior of the state machines. That means that all outputs are determined in the same cycle. Then, the communication occurs via an intermediate step using a communication bus. This causes a delay. Different bus topologies are not considered. Every failure scenario, including all timing combinations, is checked and validated. The assumptions are thus substantiated. A stationary state is needed when using bounded model checking.
\\
\\
The complexity of the arbitration logic corresponds to the magnitude of problems in the automotive industry. However, our approach is transferable to other applications since it follows a systematic procedure as defined by ISO 26262. Nevertheless, our results are based on the verification of a specific system. The simplifications of this model will determine where else it can be used. The assumptions must be evaluated for every application since the timing behaviour might differ. In conclusion, we have developed a procedure, which can be effectively used to meet the requirements of the ISO 26262 norm in the automotive industry. The time constraints need to be evaluated for each use-case.
\section{Conclusion and Future Work}
This work presents an approach to verify a fail-operational automotive system using model checking. The approach complies with the industry standard for functional safety, ISO 26262, and increases safety case as an evidence of functional safety through a verification process of a fail-operational system. 
\\
\\
Related work addresses the verification of fail-operational systems mainly by focusing on reliability. Previous applications of formal fail-operational methods in the automotive context have only addressed problems with low complexity. Our application addresses industry-relevant problems with high complexity. 
\\
In particular, our approach includes all the necessary steps to check the model and verify the safety system in compliance with ISO 26262. The model of the arbitration logic with regard to safety critical failures, formal specifications, is presented. We used the existing literature to define the system states by consolidating the individual states of each element. We defined the specification terms consisting of preconditions and target criteria. This enabled a systematic verification of the safety criteria and defined the scope of the verification in compliance with ISO 26262. Our approach includes a method for allocating constraints to failure combinations with the operation mode to limit the model checking effort and provide evidence for functional safety in compliance with ISO 26262. Therefore, we dealt with the state space explosion problem and overcame the limitations described in the literature. We did this by limiting the failure combinations to a required subset in compliance with ISO 26262, segmenting the analysis, and using bounded model checking. The verification meets the model and formal specifications. Furthermore, we addressed the tool qualification problem which is a requirement of ISO 26262 to utilize software tools in the development of safety critical systems. By verifying an actual arbitration logic, we could clarify the requirements and more importantly, identify failures in the system. Furthermore, we were able to show the application and confirm compliance with the industry standard ISO 26262 with independent reviews.
\\
\\
In summary, our work contributes to three major aspects in fail-operation systems. First, we focused on systematic failures, not realiability analysis, during the verification of fail-operational automotive systems. Second, our approach fully complies with ISO 26262, as confirmed by external reviews. This is in contrast to related work which has only been able to partially comply with ISO 26262. Third, our work uses a highly complex use case from the industry in to check the models for fail-operational systems in the automotive industry. 
\\
\\
Future work should focus on optimizing computing time and integrating the analysis in the development process. We were able to experience that a switch to Linux based systems for example and more computing power drastically reduces the computing time. The integration of formal verification might be in a model-based analysis using Matlab, Simulink, where initial research has already been conducted. The implementation of safety requirements should be conducted stepwise, as formal verification might support all phases of the product maturity.

\ifCLASSOPTIONcompsoc
  \section*{Acknowledgments}
\else
  \section*{Acknowledgment}
\fi

The authors would like to thank all the colleagues at the Institute for Software Engineering at the University of Stuttgart and BMW, for their support regarding this work, including those who are not specifically listed as authors.

\ifCLASSOPTIONcaptionsoff
  \newpage
\fi

\begin{IEEEbiography}[{\includegraphics[width=1in,height=1.25in,clip,keepaspectratio]{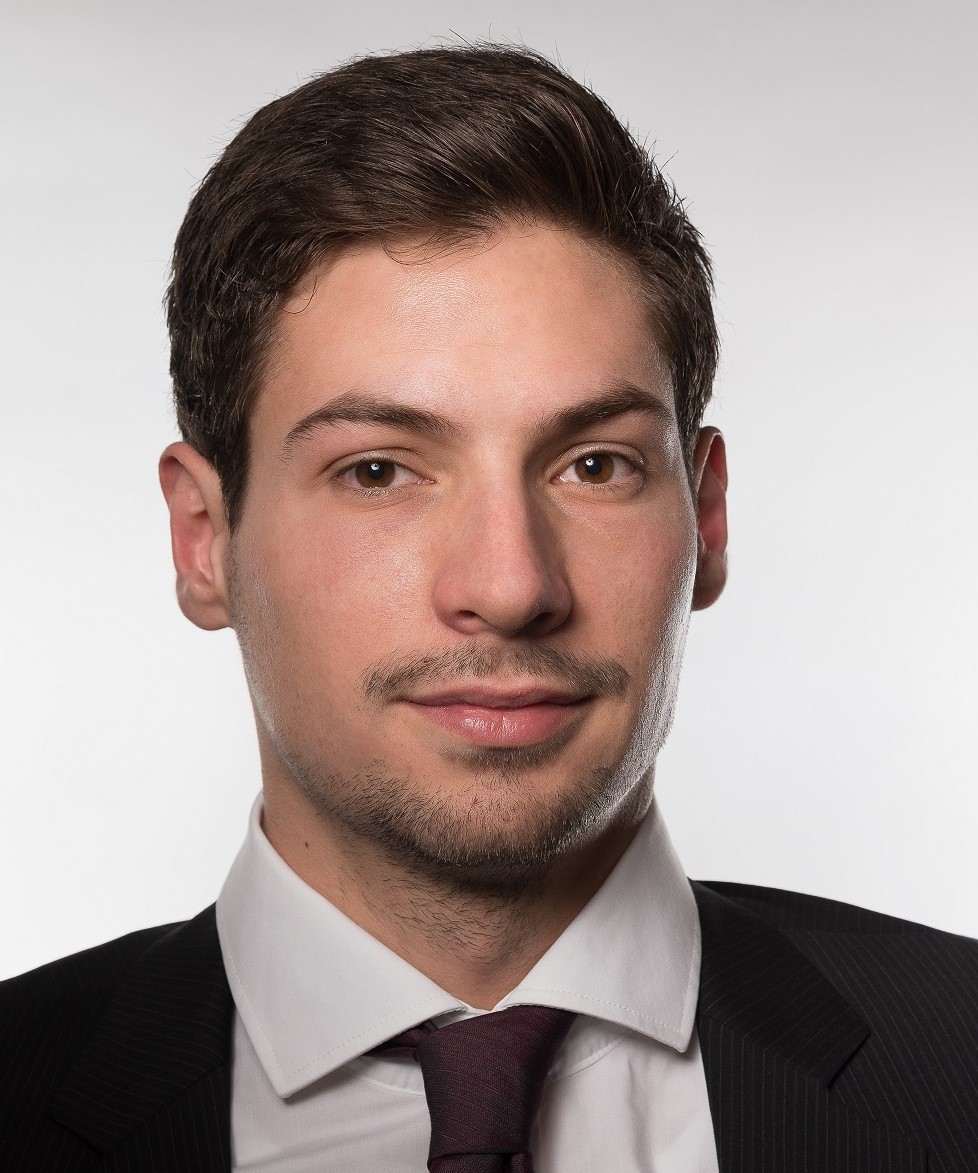}}]{Tobias Schmid}
received his Bachelor’s and Master’s degree from the Technical University of Munich in mechanical engineering. Currently he is a PhD candidate at the Institute for Software-Technology at the University of Stuttgart. He conducts research on Safety Analysis of Fail-Operational driving systems in collaboration with the BMW Group.
\end{IEEEbiography}

\begin{IEEEbiography}[{\includegraphics[width=1in,height=1.25in,clip,keepaspectratio]{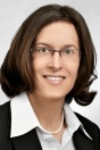}}]{Stefanie Schraufstetter}
holds a diploma in technical mathematics and a PhD in computer science both from the Technical University of Munich. After receiving her PhD, she joined the BMW Group where she is currently working as an R\& D Engineer in the field of highly autonomous driving and functional safety.
\end{IEEEbiography}

\begin{IEEEbiography}[{\includegraphics[width=1in,height=1.25in,clip,keepaspectratio]{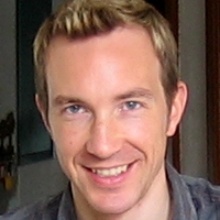}}]{Jonas Fritzsch}
Jonas Fritzsch, M.Sc. researches on Software Engineering and Architectures at the University of Stuttgart. He gathered over ten years of experience in Enterprise Software Development while working for HPE (former HP). As a university lecturer he teaches programming and algorithms to computer science students.
\end{IEEEbiography}

\begin{IEEEbiography}[{\includegraphics[width=1in,height=1.25in,clip,keepaspectratio]{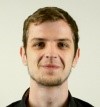}}]{Dominik Hellhake}
received his master’s degree from the Technical University of Munich in software engineering. He is a PhD candidate at the Institute for Software-Technology at the University of Stuttgart. His research is about a systematic approach for integration testing of distributed software systems in the context of automotive series development.  
\end{IEEEbiography}

\begin{IEEEbiography}[{\includegraphics[width=1in,height=1.25in,clip,keepaspectratio]{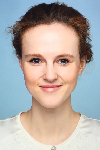}}]{Greta Koelln}
received her master’s degree from the Otto von Guericke University Magdeburg in mechanical engineering. She is a PhD candidate at the Department of Mechanical Engineering at the University Magdeburg in collaboration with the BMW Group. She conducts research in the field of Safety Analysis with a focus on System Theoretic Process Analysis in the field of automated/autonomous vehicles.
\end{IEEEbiography}

\begin{IEEEbiography}[{\includegraphics[width=1in,height=1.25in,clip,keepaspectratio]{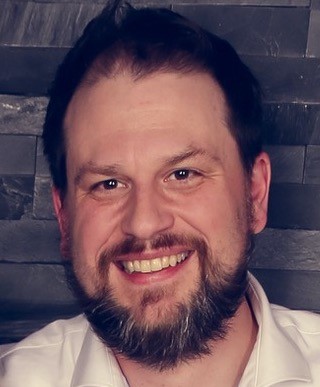}}]{Stefan Wagner}
is a full professor of empirical software engineering and managing director of the Institute of Software Engineering at the University of Stuttgart, Germany. He studied computer science in Augsburg and Edinburgh and received a PhD in software engineering from TU Munich. His research interests include software quality, requirements engineering, safety/security engineering and agile/continuous software development. He is a member of IEEE, ACM and the German GI.
\end{IEEEbiography}





\end{document}